# Transmission-matrix Quantitative Phase Profilometry for Accurate and Fast Thickness Mapping of 2D Materials

Yujie Nie[1,†], Nansen Zhou[1,†], Li Tao[2,3], Jinlong Zhu[4], Zhaoli Gao[1,5], Jianbin Xu[3], and Renjie Zhou[1,*]

[1]Department of Biomedical Engineering, The Chinese University of Hong Kong, Shatin, New Territories, Hong Kong, China
[2]Key Lab of Advanced Optoelectronic Quantum Architecture and Measurement (MOE), School of Physics, Beijing Institute of Technology, Beijing 100081, China
[3]Department of Electronic Engineering, The Chinese University of Hong Kong, Shatin, New Territories, Hong Kong, China
[4]State Key Laboratory of Digital Manufacturing Equipment and Technology, Huazhong University of Science and Technology, Wuhan 430074, China
[5]CUHK Shenzhen Research Institute, No.10, Yuexing Second Road, Nanshan, Shenzhen 518057, China

[†] These authors contributed equally to this work
*Corresponding author: rjzhou@cuhk.edu.hk

**ABSTRACT**

The physical properties of two-dimensional (2D) materials may drastically vary with their thickness profiles. Current thickness profiling methods for 2D material (e.g., atomic force microscopy and ellipsometry) are limited in measurement throughput and accuracy. Here we present a novel high-speed and high-precision thickness profiling method, termed Transmission-Matrix Quantitative Phase Profilometry (TM-QPP). In TM-QPP, picometer-level optical pathlength sensitivity is enabled by extending the photon shot-noise limit of a high sensitivity common-path interferometric microscopy technique, while accurate thickness determination is realized by developing a transmission-matrix model that accounts for multiple refractions and reflections of light at sample interfaces. Using TM-QPP, the exact thickness profiles of monolayer and few-layered 2D materials (e.g., $MoS_2$, $MoSe_2$ and $WSe_2$) are mapped over a wide field of view within seconds in a contact-free manner. Notably, TM-QPP is also capable of spatially resolving the number of layers of few-layered 2D materials.

**Keywords:** interferometric microscopy, quantitative phase microscopy, thickness profiling, 2D materials

**INTRODUCTION**

Two-dimensional (2D) materials, such as graphene [1] and transition-metal dichalcogenides (TMDs) [2], are increasingly playing critical roles in electronic and optoelectronic devices (e.g., thin-film transistors, solar cells, and light-emitting diodes) [3, 4, 5, 6]. As the properties of 2D materials are strongly dependent on their thickness and the number of layers, mapping their thickness profiles is critical for characterizing 2D material-based devices [7]. In recent years, the logic units in electronic devices are moving into sub-10 nm scale and a high production yield needs to be maintained [8], thus greatly demanding metrology tools with high accuracy and high measurement throughput [9]. A variety of metrology tools have been applied to characterizing thin-film structures including 2D materials, such as atomic force microscopy (AFM) [10], scanning electron microscopy (SEM) [11], Raman spectroscopy [12], and ellipsometry [13]. Although AFM and SEM can provide high spatial resolution and measurement precision, they pose a potential damage to a specimen and have an extremely low measurement throughput [14, 15]. Raman spectroscopy/microscopy can be applied to distinguishing the number of layers of 2D materials [16, 17] of limited types (e.g., multi-layers of graphene sheets can hardly be differentiated [18]), whereas the measurement speed is extremely low, especially when providing spatial mapping based on scanning. Imaging ellipsometry provides contact-free mapping of 2D material thickness profiles, but the involved scanning of wavelengths or illumination angles and the high complexity in model fitting can potentially hinder the measurement throughput [19].

Quantitative phase microscopy (QPM), as a highly sensitive wide-field imaging method that maps the sample optical pathlength difference (OPD), has been increasingly applied to nanometrology in recent years [20, 21, 22]. However, the OPD or phase sensitivity of current QPM methods is mostly limited to ~1 nm [23], which is insufficient for profiling single atomic layer 2D materials. To improve the phase sensitivity to ~ 0.1 nm or less, common-path interferometry and/or broadband illumination have been implemented in QPM, such as epi-illumination gradient light interference microscopy (GLIM) [24, 25], spatial light interference microscopy (SLIM) [26, 27], quadri-wave lateral shearing interferometry (QLSI) [15], and white-light diffraction phase microscopy (wDPM) [28]. These high sensitivity QPM methods are promising candidates for characterizing monolayer atomic structures, but their capabilities are limited by several key issues: (i) the broadband illumination and associated coherence issues may result in inaccurate phase and thickness estimation [29, 30]; (ii) negligence of multiple light refractions and reflection at interfaces may lead to significant errors in thickness estimations, especially in 2D materials; and (iii) further improvement on phase sensitivity and measurement throughput are limited by the interferometry design and the detection scheme.

Here, we propose Transmission-Matrix Quantitative Phase Profilometry (TM-QPP) for contact-free and wide-field thickness profiling of 2D materials with high sensitivity and high accuracy. In TM-QPP, a laser-illumination off-axis interferometry-based common-path QPM system allows us to achieve shot-noise limited phase imaging. A high sensitivity detection scheme, leveraging a high speed and high full-well-capacity camera and a frame summing method, is developed to further scale down the phase sensitivity to less than 6 picometers in time and 19 picometers in space. To precisely determine thickness from phase at each sample position, a transmission-matrix based thickness retrieval model is developed to account for multiple light refractions and reflections at sample interfaces. Using TM-QPP, not only the thickness profiles of various monolayer and few-layered 2D materials (e.g., $MoS_2$, $MoSe_2$, $WSe_2$, and graphene) can be precisely mapped, but also the number of layers can be determined. Our results are consistent with AFM and Raman spectroscopy, thus further demonstrating the great potential of TM-QPP for a wider use in the future for high-precision and high-speed thickness profiling of thin film structures.

# RESULTS

**Working principle of transmission-matrix quantitative phase profilometry (TM-QPP)**

As illustrated in Fig. 1(a), a collimated 532 nm laser beam illuminates a sample (i.e., a 2D material structure placed on a substrate) at normal incidence. Multiple refractions and reflections will be generated by the sample and substrate interfaces, as illustrated in Fig. 1(b). The total transmitted field is collected by an objective lens (OL) and collimated by a tube lens (TL). A common-path interferometer, implementing the diffraction phase microscopy (DPM) design [31] as shown in Fig. 1(d), is used to achieve high sensitivity phase measurements. The DPM part consists of a transmission grating placed at the intermediate image plane, which is followed by a 4$f$ system that contains a pinhole filter at the Fourier plane to derive a reference field and a sample field to create interferograms at the final image plane (refer to more details in **Supplementary Information, Section 1**). In DPM, the temporal phase sensitivity $\delta\phi_t$ is limited by the intrinsic photon shot-noise, other than the environmental fluctuations [32]. $\delta\phi_t$ follows the $1/\sqrt{N}$ relation where $N$ is the effective full-well-capacity (EFWC) [33]. To further scale down the phase sensitivity, a specially designed high full-well-capacity (HFWC) camera (Q-2A750m/CXP-6, Adimec; full-well-capacity of 2 million electrons) is implemented, as illustrated in Fig. 1(e). From the histogram of the interferogram (note that intensity is converted to number of electrons), we can see that the increase of EFWC improves the fringe contrast, which is the basis for achieving a high signal-to-noise ratio (SNR) in the phase image for discerning small thickness variations. However, an accurate determination of 2D material thickness map from the phase image is not straightforward; it requires carefully accounting for the multiple refractions and reflections of light at each sample and substrate interface. Therefore, a transmission-matrix model is developed to retrieve the thickness map, as illustrated in Fig. 1(c). In the following section, we introduce the transmission-matrix model.

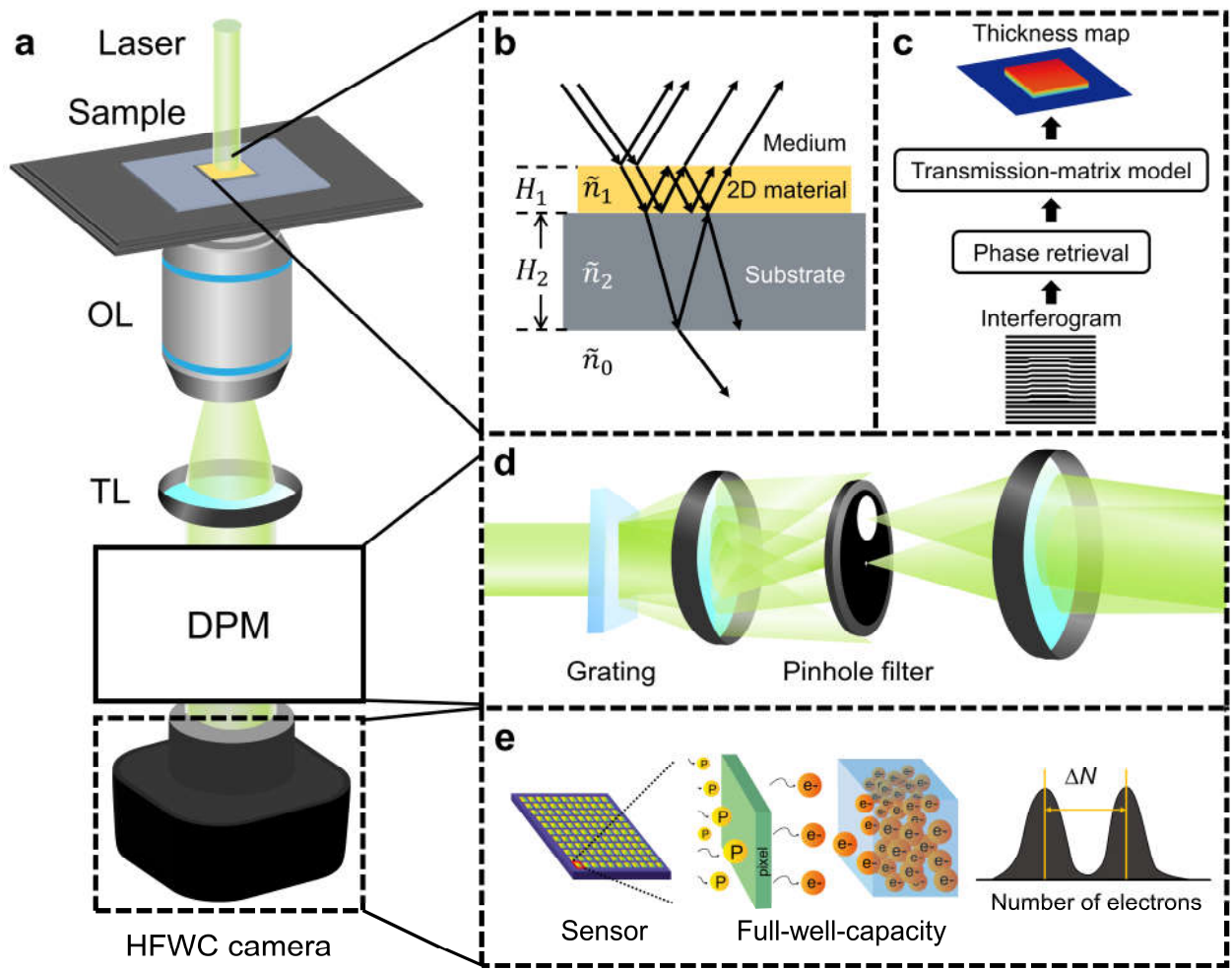

**Fig. 1. Working principle of transmission-matrix quantitative phase profilometry.** (a) Schematic design of the system. OL: objective lens; TL: tube lens; HFWC camera: high full-well-capacity camera. (b) Description of multiple refractions and reflections at sample and substrate interfaces. $H_1$ and $H_2$ represent the thicknesses of the 2D material and the substrate, respectively. $\tilde{n}_0$, $\tilde{n}_1$, and $\tilde{n}_2$ represent the refractive index values of the medium, the 2D material, and the substrate, respectively. (c) Illustration of sample thickness map converted from the interferogram using the transmission-matrix model. (d) Design of the DPM-based common-path interferometer. (e) Illustration of scaling down phase sensitivity by extending the fringe contrast through increasing the effective full-well-capacity.

**Transmission-matrix model for thickness retrieval**

The sample thickness profile is usually retrieved from the measured phase map $\Delta\varphi_m(x,y)$ using a simple linear model: $H_{1m}(x,y) = \dfrac{\Delta\varphi_m(x,y) \cdot \lambda_0}{2\pi \cdot Re(\Delta\tilde{n})}$, where $\lambda_0$ is the central wavelength of the illumination source; and $Re(\Delta\tilde{n})$ is the real part of refractive index difference between the sample and the medium ($Re(\Delta\tilde{n})$ is usually assumed to be small) [34]. However, directly applying this model to 2D materials can lead to large thickness estimation errors due to the following reasons: (i) $\Delta\tilde{n}$ could be large that leads to multiple reflections and refractions at the sample and substrate interfaces, as illustrated in Fig. 1(b); and (ii) the neglected imaginary parts of the complex refractive index (or the extinction coefficients) may result in phase shifts at interfaces. Therefore, we propose to develop a transmission-matrix model to accurately describe light propagation in the sample and substrate as illustrated in Fig. 2(a), and subsequently establish a relationship among incident field $E_{in}$, reflected field $E_r$, and transmitted field $E_t$ as:

$$\begin{bmatrix} E_{in} \\ E_r \end{bmatrix} = M \begin{bmatrix} E_t \\ 0 \end{bmatrix}, \quad (1)$$

where $M = D_{0,1} P_1 D_{1,2} P_2 ... D_{N-1,N} P_N D_{N,0}$ is the total transmission matrix with $D$ and $P$ represent the transmission matrix at each interface and the propagation matrix in each sample layer, respectively. When considering just a piece of 2D material placed on a substrate, then $M = D_{0,1} P_1 D_{1,2} P_2 D_{2,0}$, where $D_{2,0}$, $P_2$, $D_{1,2}$, $P_1$ and $D_{0,1}$ represent the matrices at medium and substrate interface, substrate, substrate and 2D material interface, 2D material, and 2D material and medium interface, respectively. By applying the transmission-matrix model, we link $E_t$ with the complex refractive index and thickness of 2D material structure ($\tilde{n}_1, H_1$), substrate ($\tilde{n}_2, H_2$), and the medium ($\tilde{n}_0$). It should be noted that it is crucial to consider the substrate thickness $H_2$ and refractive index $\tilde{n}_2$ to ensure an accurate thickness retrieval, but this was typically ignored in reported methods in literature. It should also be noted that for 2D materials with multiple atomic layers, their refractive index values vary with the number of layers [35], which will also be accounted for by our model.

Using the transmission-matrix model, we first simulated retrieved thickness $H_{1s}$ from the linear model and calculated the thickness error rate as $R_e = |H_{1s} - H_1|/H_1$ on different 2D materials, as shown in Fig. 2(b). The refractive index values of the 2D materials corresponding to different atomic layers are used, including molybdenum diselenide (MoSe$_2$) of 1-3 layers [36], Tungsten diselenide (WSe$_2$) of 1-4 layers [37], and monolayer molybdenum disulfide (MoS$_2$) [38]. A sapphire substrate with a thickness of $H_2 = 260 \mu m$ is assumed. Our results show that for mono- or few-layered 2D materials (<5 layers, i.e., light gray region in Fig. 2(b)), the error rate is around or over 100%, thus it is critical to correct their thickness errors. As the number of layer increases (>5 layers), the 2D materials become bulk materials [39]. As expected, the error rate is asymptotically approaching 0%, as exemplified on MoSe$_2$ [36] (refer to MoSe$_2$ bulk curve in Fig. 2(b)). Note that the error rate simulation is based on available complex refractive index values at the central wavelength of 532 nm from the cited references. A table summarizing the refractive index values of all used 2D materials and substrate types is provided in **Supplementary Section 7**. More details on the error rate analysis between linear and transmission-matrix models and experimental verifications are included in **Supplementary Information, Section 2.1-2.4.** In our experiments, $E_t$ can be retrieved from the TM-QPP system, from which the measured linear model thickness $H_{1m}$ can be directly obtained. Using the transmission-matrix model, the corrected thickness $H_{1c}$ can be obtained from $H_{1m}$ with a correction term $\Delta\phi$ through solving the following transcendental equation:

$$f(H_{1c}) = H_{1c} + \Delta\phi\{H_{1c}; \tilde{n}_0, \tilde{n}_1, \tilde{n}_2, H_2\}/(k_1 - k_0) - H_{1m} = 0, \qquad (2)$$

where $k_{0,1} = 2\pi\tilde{n}_{0,1}/\lambda_0$ and $\tilde{n}_0$, $\tilde{n}_1$, $\tilde{n}_2$, and $H_2$ are known parameters. More details on the transmission-matrix model derivation and the related thickness retrieval algorithm are included in **Supplementary Information, Section 2.2.**

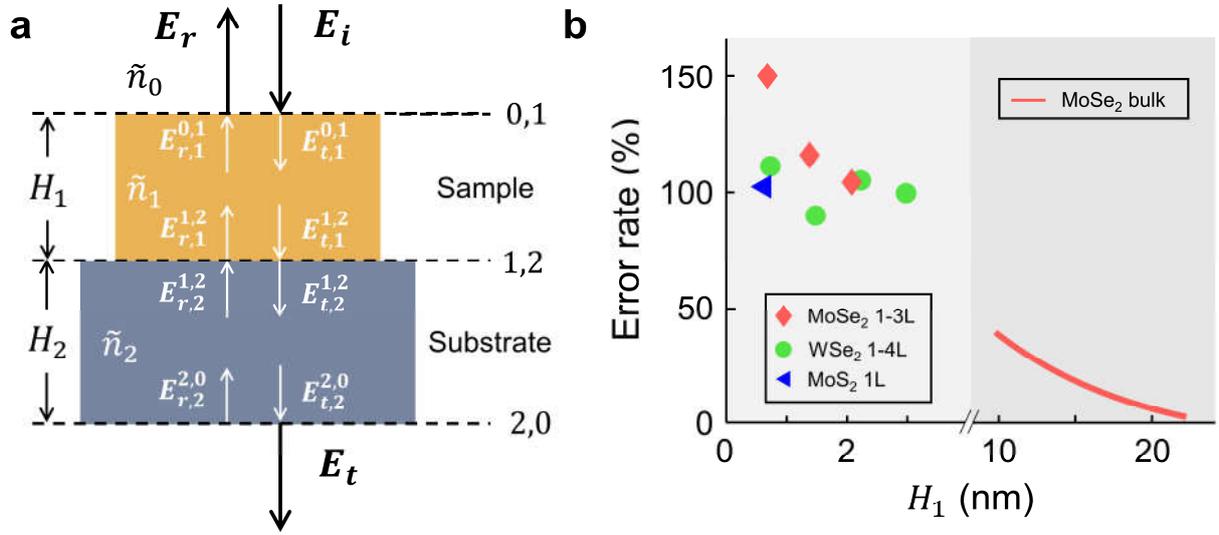

**Fig. 2. Transmission-matrix model for correcting thickness retrieval of 2D materials.** (a) Description of light propagation in the sample and substrate layers. $E_i$, $E_r$ and $E_t$ represent the incident field, the reflected field, and the transmitted field, respectively; $\tilde{n}_0$, $\tilde{n}_1$, and $\tilde{n}_2$ represent the refractive index values of the surrounding medium, sample (i.e., 2D material), and substrate, respectively; $H_1$ and $H_2$ represent the thickness of sample and substrate, respectively; and $E_{r,j}^{j,j+1}$ and $E_{t,j}^{j,j+1}$ ($j=1,2$) represent the reflected and transmitted fields at sample and substrate layers and their interfaces. (b) Thickness error rate simulation on MoSe$_2$, WSe$_2$, and MoS$_2$ of different numbers of layers and bulk MoSe$_2$.

**Realization of high sensitivity phase retrieval**

Under shot-noise limit, the temporal phase sensitivity of our TM-QPP system follows the relation $\delta\phi_t \approx \frac{1}{\sqrt{m \cdot N}}$ [33], where $N$ is the EFWC that can be estimated from the histogram of the interferogram as illustrated in Fig. 1 (e), while $m$ is the effective number of pixels in the diffraction spot that is calculated to be around 11 based on our system configuration. To characterize the phase sensitivity of our system, we measured time-series of interferograms at a fixed frame rate of 500 frames per second (fps) with an exposure time of 1936 μs and calculated the temporal phase sensitivity in the form of OPD as $OPD_t = \frac{\lambda_0 \cdot \delta\phi_t}{2\pi}$. $OPD_t$ is estimated from each time-series that contains 600 interferograms and averaged over a total of 15 different time-series, following a similar procedure in Ref. [33]. By varying the illumination intensity, EFWC was changed from 100,000 to 600,000 e$^-$ and $OPD_t$ vs. EFWC is plotted in Fig. 3(a) (red curve). The $OPD_t$ curve shows a clear scaling down as EFWC increases, which matches with the theoretical curve (black curve). A larger discrepancy between the experiment and theory at the lower EFWC end is likely due to the relatively large read noise of 969 e$^-$ of the HFWC camera [40]. In our experiment, the best $OPD_t$ achieved is around 33 picometers (pm) when EFWC is at around 700,000 e$^-$. Note that for conventional cameras, the FWC is on the order of 10,000 e$^-$ (e.g., when using a PCO camera, model pco.edge 4.2LT, we obtained EFWC of around 13,000 e$^-$), resulting in a $OPD_t$ of around 0.23 nm. We also characterized the spatial phase sensitivity $OPD_s$ as shown in the green curve. $OPD_s$ is related to the uniformity of the temporal phase sensitivity across the whole field of view.

To further scale down the phase sensitivity, a frame summing method is developed as illustrated in Fig. 3(b). We initially sequentially measure a stack of 600 interferograms and divide them into $S$ groups with each group containing $p$ frames, i.e., $p \times S = 600$. Within each group, $p$ frames are summed together pixel by pixel to generate a summed interferogram as illustrated in Fig. 3(b), where we show the first summed frame in B as marked by a red rectangular box. The first summed frame is used for calibrating the retrieved phase maps corresponding to the subsequent $S-1$ summed interferograms. In Fig. 3(c), we summarize the $OPD_t$ and $OPD_s$ values computed with different $p$ values in a table, where we can see the enhancement of both spatial and temporal phase sensitivities. When $p=200$ and $S=3$ (note that $S$ should be at least 3 to calculate $OPD_t$), we achieved $OPD_t$ of 5.9 pm, which is improved from the 32.7 pm when frame summing is not applied (i.e., $p=1$ and $S=600$). At the same time, $OPD_s$ is improved to 18.1 pm from 47.5 pm. As the thickness of single-layered 2D materials is larger than 0.1 nm, summing 50 frames to achieve $OPD_t$ of 9.4 pm and $OPD_s$ of 27.5 pm is sufficient for profiling 2D materials. Summing more frames will prolong both the measurement time and the data processing time. Therefore, we keep $p=50$ for later experiments. Figure 3(d) and (e) show the representative OPD maps with and without applying 50-frame summing, respectively. The standard deviation of the OPD map is reported as the $OPD_s$ value.

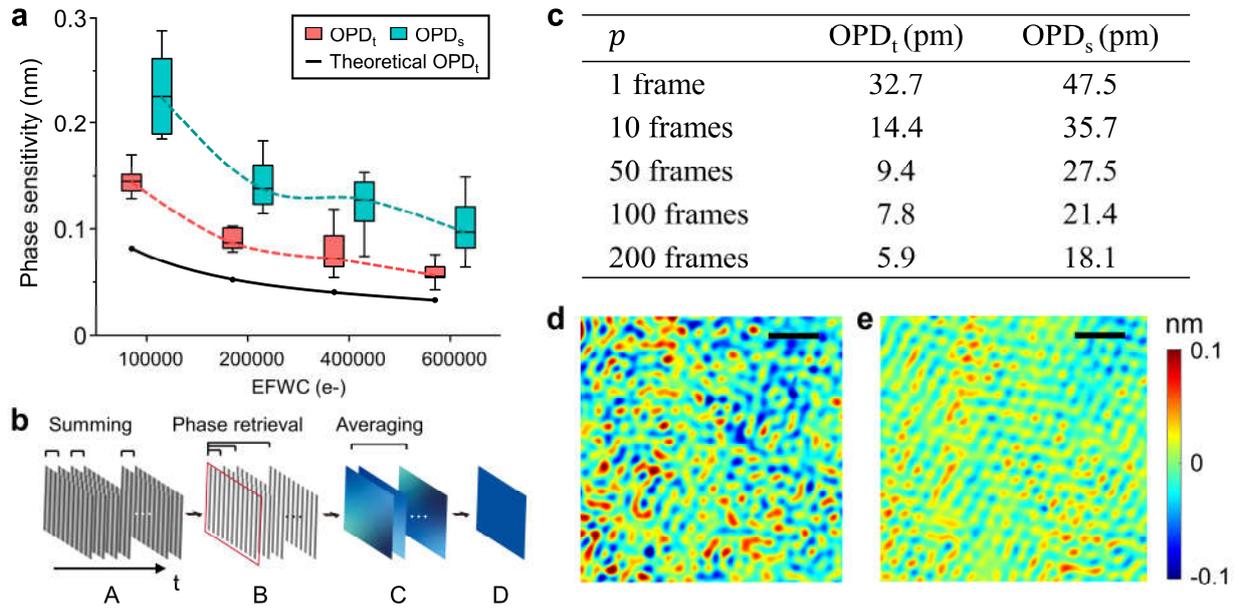

**Fig. 3. Experimental demonstration of achieving high phase sensitivity.** (a) Phase sensitivity vs. EFWC. The green, red, and black curves represent the measured spatial phase sensitivity ($OPD_s$), measured temporal phase sensitivity ($OPD_t$) and theoretical temporal phase sensitivity (theoretical $OPD_t$) respectively. (b) Illustration of the frame summing method for further scaling down the phase sensitivity. A. Summing of frames or interferograms. B. Phase retrieval. C. Averaging of phase maps. D. Resulting phase map. (c) Phase sensitivity improvement through frame summing. (d,e) Spatial phase noise maps with and without applying the 50-frame summing, respectively. Scale bar: 5 μm.

**Thickness profiling of monolayer 2D material structures**

To verify the accuracy of the transmission-matrix model in thickness profiling of 2D materials, we first tested on a monolayer MoS$_2$ sample fabricated by chemical vapor deposition (CVD) and transferred to a sapphire substrate. The bright-field optical image in Fig. 4(a) shows triangular MoS$_2$ flakes that have side lengths vary from approximately 5 to 12 μm. The atomic structure of monolayer MoS$_2$ is illustrated in the inset of Fig. 4(a). A unit cell is composed of one molybdenum atom and two sulfur atoms that are arranged in a sandwich structure by the S-Mo-S covalent bonds. To verify that the MoS$_2$ flakes are single-layered structures, we first used a commercial AFM to scan the selected region as indicated in Fig. 4(b). A line profile is shown in the inset of Fig. 4(b), where a difference of approximately 0.7 nm is measured. To further verify the structures, we also measured the Raman and photoluminescence (PL) spectra. Typically, two Raman peaks, $E^1_{2g}$ and $A_{1g}$, can be observed in the Raman spectrum that reflect the crystal structures of MoS$_2$. Figure 4(c) shows two distinguished Raman peaks ($E^1_{2g} \approx 386.5$ cm$^{-1}$, $A_{1g} \approx 405.5$ cm$^{-1}$) with a frequency difference of around 19 cm$^{-1}$, which agree with previous reports [41]. A Raman peak at approximately 418 cm$^{-1}$ is also detected, which corresponds to the sapphire substrate [42]. A prominent PL peak at 673 nm is observed as shown in Fig. 4(d), thus reaffirming the flakes are single-layered [43].

Figure 4(e) shows the thickness map of the MoS$_2$ flakes obtained from our TM-QPP system with the 50-frame summing method. In the transmission-matrix model, we used refractive index of 4.4080+0.6290$i$ for monolayer MoS$_2$ as reported in [38] and refractive index of 1.7717 for the sapphire substrate (thickness of 260 μm) as reported in [44]. The zoom-in thickness map of one flake, enclosed in the black box, is shown in Fig. 4(f), where a line profile is plotted in the inset. From the histogram in Fig. 4(g) that is fitted with a nonparametric kernel-smoothing distribution (black solid line), a thickness of around 0.65 nm is determined, which is in a close agreement with the AFM measurement and the reported value of monolayer MoS$_2$ [45], but over 100 times faster than the AFM. For example, scanning a 10×10 μm$^2$ area with AFM (Bruker Dimension ICON) needs ~ 320 sec with scan rate at 0.8 Hz, 256 lines, while our TM-QPP system only needs ~ 2 sec for single frame analysis (note that the image acquisition time is only a few milliseconds) and ~ 5 sec when 50-frame summing is applied based on 150 measurements (note that the thickness retrieval process is performed using MATLAB on Core$^{TM}$ i7-6850K CPU). An accurate thickness profiling of the monolayer MoS$_2$ can also be obtained without frame summing as shown in **Fig. S5**, in which we also verified the high measurement stability of our method. We noted that the refractive index values of monolayer MoS$_2$ reported in the literature are different that may greatly affect the retrieved thickness values when using our model. Therefore, we calculated the corresponding thickness maps when using several representative refractive index values and summarized our results in **Supplementary Information 3.2.** Although the refractive index values vary, our model can tolerate such variations as the fluctuation of the thickness is around 0.17 nm (i.e., 26.2 %) from 0.65 nm as shown in **Fig. S6**. A comparison of thickness profile results on the monolayer MoS$_2$ structure between the HFWC camera and a normal camera (i.e., the PCO camera) with frame-summing method is shown in **Fig. S10**. The comparison shows a great improvement in the image quality by utilizing the HFWC camera. In addition, the unevenness of the substrate presents a strong background phase variation as seen in Fig. 4(e). Therefore, although the system phase sensitivity is high enough to resolve monolayer materials, the ultimate measurement limit may lie in the flatness of the substate. This becomes more obvious when imaging thinner monolayer structures, e.g., graphene. The thickness profiling result of a monolayer graphene sample using our system is presented in **Supplementary Information 4**. We are able to obtain the thickness of the graphene sample, but the image quality is not high which is mostly attributed to the unevenness of the substrate.

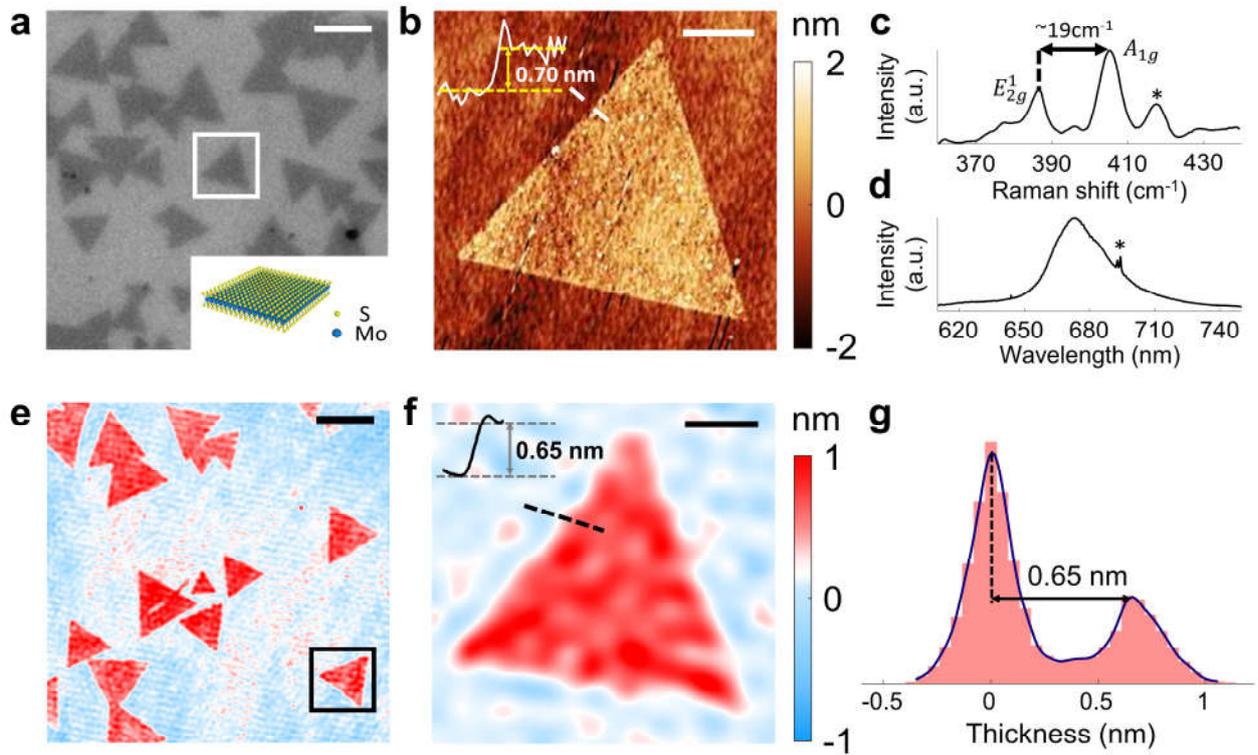

**Fig. 4. Thickness profiling of monolayer MoS$_2$ structures.** (a) Bright-field optical image of the fabricated monolayer MoS$_2$ flakes on a sapphire substrate. Inset shows the atomic structure of monolayer MoS$_2$. Scale bar: 10 μm. (b) AFM image of the selected flake in (a). Scale bar: 2 μm. (c)&(d) Raman and PL spectra of (b), respectively. The star symbols represent the characteristic peaks of sapphire substrate. (e) The thickness map of the MoS$_2$ flakes retrieved using TM-QPP with 50-frame summing. Scale bar: 10 μm. (f) The zoom-in of the flake enclosed in the black box in (e). Scale bar: 2 μm. (g) Histogram of the thickness map in (f).

**Differentiating the number of layers of few-layered 2D material structures**

Two or more layers of 2D materials can form few-layered 2D materials in a certain stacking order [46]. The change of thickness can significantly affect the electronic structures of 2D materials, e.g., transition from an indirect bandgap in the bulk to a direct gap in the monolayer [47]. Therefore, it is critical to determine the number of layers of 2D materials. Bright-field light microscopy based methods have been applied to identify the thickness of multilayers by analyzing the contrast difference between 2D materials and their substrates [48, 49], but they do not provide quantitative measurements of thickness and the contrast could be largely affected by the illumination source, the intensity balance, and etc. Rayleigh scattering [50], Raman spectroscopy [51], and second harmonic generation [52], can potentially distinguish different layers of certain types of 2D materials, but they lack the spatial information and quantitative measure of thickness values. Therefore, a method that can directly map the exact thickness of few-layered 2D materials and differentiate their thicknesses without limitations on material types is yet to be developed.

To explore the capability of TM-QPP for differentiating the number of layers of 2D materials, we used few-layered $WSe_2$ and $MoSe_2$ samples with different numbers of layers. Each sample is scattered on a 260-μm thick sapphire substrate. From bright-field optical images, the layered structures can be distinguished owing to the intensity contrast, as shown in the Fig. 5(a) and (e). Using our system, we first obtained the high sensitivity phase maps of $WSe_2$ and $MoSe_2$ in Fig. 5(b) and (f). Combining with the bright-field optical images, we selected several regions in the sample area (region ii, iii, and iv) for subsequent quantification of the number of layers and a background region (region i) for calibration (i.e., thickness set as zero). Noted that as the refractive index values in different regions of the sample could be different due to the variation of the layer numbers, the phase maps cannot be directly converted into exact thickness maps in one single step. In the first step, we used the refractive index values of monolayer $WSe_2$ and monolayer $MoSe_2$ to initialize the thickness estimation for each region when using our transmission-matrix model. In the second step, we estimated the number of layers $L_N$ of each region through $L_N = H_{1c} / h$, where $H_{1c}$ being the peak location of the thickness histogram of that region (**Fig. S9**), and $h$ being the thickness of the monolayer sample. According to the literature reported values, we set $h$ to be 0.75 nm [37] for $WSe_2$ and 0.7 nm [53] for $MoSe_2$, respectively. In the third step, with the assumption that the thickness of *l*- layered sample is $H_1 = l \cdot h$ ($l$ = 1, 2, 3, 4, ...) , we calculate the integer number $L_N$ closest to $(H_1/h)$ for each region. If $L_N$ is consistent with the assumed number of layers $l$, then the corresponding region is confirmed to be $l$-layered. The procedure will be repeated until the layer number of all regions are determined. (A detailed procedure in determining the layer number is described in **Supplementary Section 5.2**; the refractive index values of WSe$_2$ with 1-4 layers and MoSe$_2$ with 1-3 layers are provided in **Supplementary Section 7 Table S1**). Figure 5(c) and (g) show the final determined layer numbers in region ii-iv of WSe$_2$ (labeled as 1L, 2L, 4L) and MoSe$_2$ (labeled as 2L, 3L, and 4L) and the corresponding thickness values (error bar is determined by the standard deviation of the corresponding region). For WSe$_2$, region ii: 0.893±0.443 nm (1L), region iii: 1.517±0.229 nm (2L), region iv: 3.053±0.509 nm (4L); for MoSe2, region ii: 1.541±0.424 nm (2L), region iii: 2.111±0.408 nm (3L), region iv: 3.030±0.342 nm (4L). Reference thickness values $l \cdot h$ for the corresponding layer numbers ($l$) are also marked (dashed lines in Fig. 5(c) and (g)). We also used AFM to map the same regions, as shown in Fig. 5(d) and (h). In the AFM images, line profiles along the yellow lines are plotted for discerning the number of layers. However, based on the AFM images, the layer numbers cannot be easily resolved, especially when the layer number difference is 1. The Raman spectra (**Fig. S8**) are measured in each region to confirm the structures are few-layered, although the exact number of layers cannot be determined.

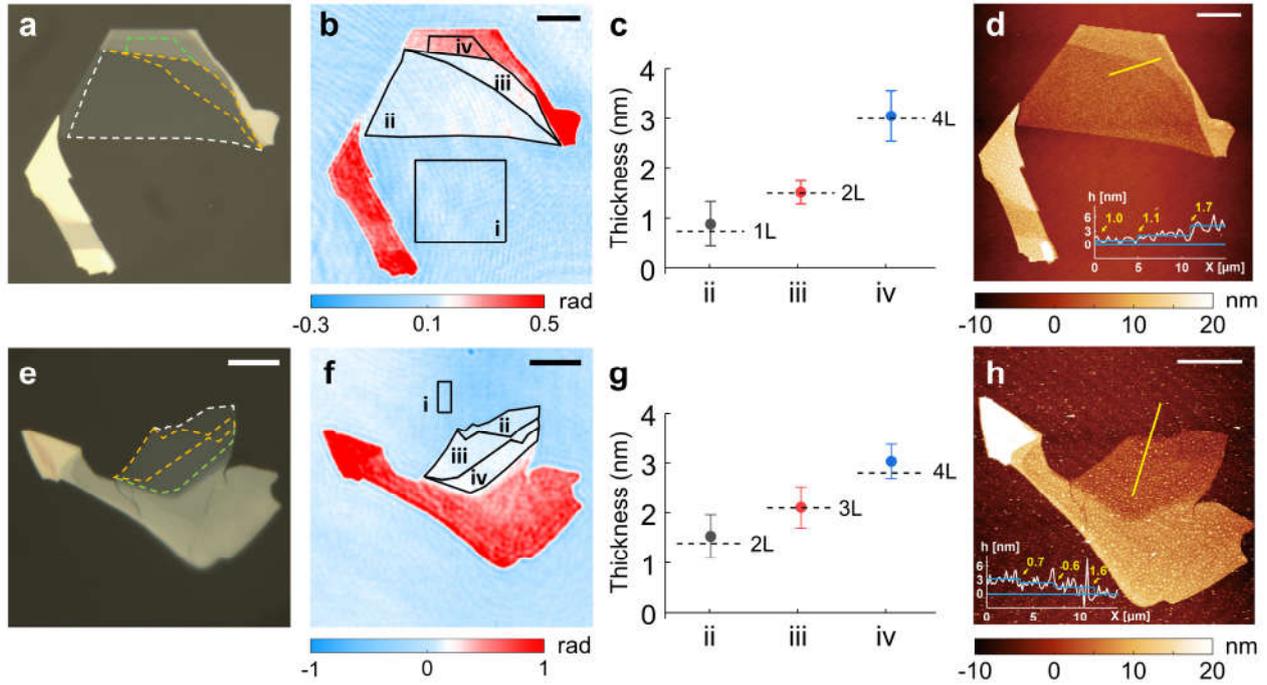

**Fig. 5. Differentiating the number of layers of few-layered WSe$_2$ and MoSe$_2$ structures.** (a)&(e) Bright-field optical images of mechanically exfoliated WSe$_2$ and MoSe$_2$ flakes on sapphire substrate, respectively. Dashed lines with different colors separate different layer regions of the flakes. (b)&(f) Corresponding phase maps of the WSe$_2$ and MoSe$_2$ flakes. (c)&(g) Thickness values calculated for each region with determined layer numbers. Dashed lines represent the reference thickness values for the corresponding layer numbers. (d)&(h) AFM images of the corresponding WSe$_2$ and MoSe$_2$ flakes. Scale bar: 10 μm.

## DISCUSSION

We demonstrated TM-QPP as a new method for fast and accurate thickness profiling of 2D materials by combining a high sensitivity QPM method with a transmission-matrix model that accounts for the multiple reflection and refractions in the sample region. The presented approach can achieve thickness profile retrieval with a temporal sensitivity of <6 pm and spatial sensitivity of <19 pm with 200-frame summing. Compared with most currently used methods (e.g., AFM and Raman spectroscopy/microscopy that are based on point-scanning or line-scanning), our method is much more efficient with 100 times higher measurement throughput and can also determine the layer number of few-layered 2D materials.

In the future, to further improve the performance of our method, e.g., profiling 2D materials with low refractive index contrast, the spatial phase sensitivity needs to be improved. The spatial sensitivity is limited by the speckle noise from the laser, which can be reduced by using a low coherence light source, such as light-emitting diode, super-luminescent diode or supercontinuum laser. Meanwhile, using substrates of lower roughness can also improve the phase image quality in experiments. Our method can also be extended to 2D materials or thin film structures on reflective substrates (e.g., silicon substrates) or both transmissive and reflective surfaces by adding a reflection mode. We envision this new wide-field optical imaging-based non-contact thickness profiling tool will find applications in high throughput and in-situ inspection of nanomaterials.

## Materials and Methods

### Raman imaging and PL spectra

The Raman and PL spectra measurements were performed on Horiba LabRAM HR Evolution Raman spectrometer with an excitation wavelength of 532 nm and power of 1 mW. The size of laser spot was ~1 μm.

### AFM imaging

The AFM measurements were conducted on Bruker Dimension ICON (tapping mode), while the quantitative analysis of the image profiles was carried out using Gwyddion software [54].

### Optical imaging

The bright-field optical images were captured on Zeiss Axio Observer 7.

### Sample preparation

Monolayer $MoS_2$, monolayer graphene, multilayer $WSe_2$ samples were purchased from Six-carbon technology, Shenzhen, China. Multilayer $MoSe_2$ sample was purchased from Beike 2D materials Co. Ltd.

### Data availability

The datasets generated during the current study are available from the corresponding author upon reasonable request.

### Code availability

The code used for simulation and data plotting is available from the corresponding author upon reasonable request.

### Acknowledgements

R. Z. acknowledges the financial support from Hong Kong Innovation and Technology Fund (Grant Nos. ITS/148/20 and ITS/178/20FP), Croucher Foundation (CM/CT/CF/CIA/0688/19ay), and Research Grant Council of Hong Kong SAR (Grant No. 14209521); Z. G. acknowledges the financial support from the Key-Area Research and Development Program of Guangdong Province (Grant No. 2020B0101030002), the National Natural Science Foundation of China (Grant No.

62101475), the Research Grant Council of Hong Kong (Grant Nos. 24201020 and 14207421), and the Research Matching Grant Scheme of Hong Kong Government (Grant No. 8601547). J. X. acknowledges the financial support from the Research Grants Council of Hong Kong SAR (Grant No. N_CUHK 438/18) and CUHK Group Research Scheme (Grant No. AoE/P-701/20, 14203018). J. Z. acknowledges the financial support from National Natural Science Foundation of China (Grant No. 52175509). L. T. acknowledges the financial support support from the National Natural Science Foundation of China (Grant No. 62005051).

# REFERENCES

1. Novoselov, K.S. et al. A roadmap for graphene. *Nature* **490**, 192-200 (2012).

2. Li, X. et al. Graphene and related two-dimensional materials: Structure-property relationships for electronics and optoelectronics. *Appl. Phys. Rev.* **4**, 021306 (2017).

3. Long, M., Wang, P., Fang, H. & Hu, W. Progress, challenges, and opportunities for 2D material based photodetectors. *Adv. Funct. Mater.* **29**, 1803807 (2019).

4. Tao, L. et al. Enhancing light-matter interaction in 2D materials by optical micro/nano architectures for high-performance optoelectronic devices. *InfoMat* **3**, 36-60 (2021).

5. Zhu, L. et al. Angle-selective perfect absorption with two-dimensional materials. *Light Sci. Appl.* **5**, e16052 (2016).

6. Chen, X. et al. CVD-grown monolayer $MoS_2$ in bioabsorbable electronics and biosensors. *Nat. Commun.* **9**, 1690 (2018).

7. Lemme, M. C., Li, L. -J., Palacios, T. & Schwierz, F. Two-dimensional materials for electronic applications. *Mrs. Bull* **39**, 711-718 (2014).

8. Zhu, J. et al. Optical wafer defect inspection at the 10 nm technology node and beyond. *Int. J. Extrem. Manuf.* **4**, 032001 (2022).

9. Akinwande, D. et al. Graphene and two-dimensional materials for silicon technology. *Nature* **573**, 507-518 (2019).

10. Zhu, Y. Y., Ding, G. Q., Ding, J. N. & Yuan, N. Y. AFM, SEM and TEM studies on porous anodic alumina. *Nanoscale Res. Lett.* **5**, 725 (2010).

11. Loomis, J. & Panchapakesan, B. Dimensional dependence of photomechanical response in carbon nanostructure composites: a case for carbon-based mixed-dimensional systems. *Nanotechnology* **23**, 215501 (2012).

12. Wu, J. & Xie, L. Structural quantification for graphene and related two-dimensional materials by Raman spectroscopy. *Anal. Chem.* **91**, 468-481 (2019).

13. Chang, Y. C. et al. Extracting the complex optical conductivity of mono-and bilayer graphene by ellipsometry. *Appl. Phys. Lett.* **104**, 261909 (2014).

14. Yang, W. et al. Rapid characterization of nano-scale structures in large-scale ultra-precision surfaces. *Opt. Lasers Eng.* **134**, 106200 (2020).

15. Khadir, S. et al. Optical imaging and characterization of graphene and other 2D materials using quantitative phase microscopy. *ACS Photonics* **4**, 3130-3139 (2017).

16. Ni, Z., Wang, Y., Yu, T. & Shen, Z. Raman spectroscopy and imaging of graphene. *Nano Res.* **1**, 273-291 (2008).

17. Nair, S. et al. Algorithm-improved high-speed and non-invasive confocal Raman imaging of 2D materials. *Natl. Sci. Rev.* **7**, 620-628 (2020).

18. Ni, Z. H. et al. Graphene thickness determination using reflection and contrast spectroscopy. *Nano Lett.* **7**, 2758-2763 (2007).

19. Matković, A. et al. Spectroscopic imaging ellipsometry and Fano resonance modeling of graphene. *J. Appl. Phys.* **112**, 123523 (2012).

20. Hwang, S. W. et al. Dissolution chemistry and biocompatibility of single-crystalline silicon nanomembranes and associated materials for transient electronics. *ACS Nano* **8**, 5843-5851 (2014).

21. Khadir, S. et al. Metasurface optical characterization using quadriwave lateral shearing interferometry. *ACS Photonics* **8**, 603-613 (2021).

22. Edwards, C., Arbabi, A., Popescu, G. & Goddard, L. L. Optically monitoring and controlling nanoscale topography during semiconductor etching. *Light Sci. Appl*. **1**, e30 (2012).

23. Toda, K., Tamamitsu, M. & Ideguchi, T. Adaptive dynamic range shift (ADRIFT) quantitative phase imaging. *Light Sci. Appl*. **10**, 1 (2021).

24. Kandel, M. E. et al. Epi-illumination gradient light interference microscopy for imaging opaque structures. *Nat. Commun*. **10**, 4691 (2019).

25. Nguyen, T. H. et al. Gradient light interference microscopy for 3D imaging of unlabeled specimens. *Nat. Commun*. **8**, 210 (2017).

26. Wang, Z. et al. Spatial light interference microscopy (SLIM). *Opt. Express* **19**, 1016-1026 (2011).

27. Wang, Z. et al. Topography and refractometry of nanostructures using spatial light interference microscopy. *Opt. Lett*. **35**, 208-210 (2010).

28. Bhaduri, B., Pham, H., Mir, M. & Popescu, G. Diffraction phase microscopy with white light. *Opt. Lett*. **37**, 1094-1096 (2012).

29. Nguyen, T. H., Edwards, C., Goddard, L. L. & Popescu, G. Quantitative phase imaging with partially coherent illumination. *Opt. Lett*. **39**, 5511-5514 (2014).

30. Edwards, C. et al. Effects of spatial coherence in diffraction phase microscopy. *Opt. Express* **22**, 5133-5146 (2014).

31. Popescu, G., Ikeda, T., Dasari, R. R. & Feld, M. S. Diffraction phase microscopy for quantifying cell structure and dynamics. *Opt. Lett*. **31**, 775-777 (2006).

32. Hosseini, P. et al. Pushing phase and amplitude sensitivity limits in interferometric microscopy. *Opt. Lett.* **41**, 1656-1659 (2016).

33. Nie, Y. & Zhou, R. Beating temporal phase sensitivity limit in off-axis interferometry based quantitative phase microscopy. *APL Photonics* **6**, 011302 (2021).

34. Bhaduri, B. et al. Diffraction phase microscopy: principles and applications in materials and life sciences. *Adv. Opt. Photonics* **6**, 57-119 (2014).

35. Georgescu, G. & Petris, A. Analysis of thickness influence on refractive index and absorption coefficient of zinc selenide thin films. *Opt. Express* **27**, 34803-34823 (2019).

36. Hsu, C. et al. Thickness-dependent refractive index of 1L, 2L, and 3L $MoS_2$, $MoSe_2$, $WS_2$, and $WSe_2$. *Adv. Opt. Mater*. **7**, 1900239 (2019).

37. Gu, H. et al. Layer-dependent dielectric and optical properties of centimeter-scale 2D $WSe_2$: evolution from a single layer to few layers. *Nanoscale* **11**, 22762-22771 (2019).

38. Zhang, H. et al. Measuring the refractive index of highly crystalline monolayer MoS$_2$ with high confidence. *Sci. Rep*. **5**, 8440 (2015).

39. Zhang, X., Kawai, H., Yang, J. & Goh, K. E. J. Detecting MoS$_2$ and MoSe$_2$ with optical contrast simulation. *Prog. Nat. Sci*. **29**, 667-671 (2019).

40. Meynants, G. et al. 700 frames/s 2 MPixel global shutter image sensor with 2 Me-full well charge and 12 μm pixel pitch. *Proc. IISW* 409-413 (2015).

41. Lee, K. C. et al. Plasmonic gold nanorods coverage influence on enhancement of the photoluminescence of two-dimensional MoS$_2$ monolayer. *Sci. Rep*. **5**, 16374 (2015).

42. Deng, S., Loterie, D., Konstantinou, G., Psaltis, D. & Moser, C. Raman imaging through multimode sapphire fiber. *Opt. Express* **27**, 1090-1098 (2019).

43. Wang, Z. et al. Plasmonically enhanced photoluminescence of monolayer MoS$_2$ via nanosphere lithography-templated gold metasurfaces. *Nanophotonics* **10**, 1733-1740 (2021).

44. Malitson, I. H. & Dodge, M. J. Refractive-index and birefringence of synthetic sapphire. *J. Opt. Soc. Am.* **62**, 1405 (1972).

45. Li, X. & Zhu, H. Two-dimensional MoS$_2$: Properties, preparation, and applications. *J. Materiomics* **1**, 33-44 (2015).

46. Li, X. -L. et al. Layer-number dependent optical properties of 2D materials and their application for thickness determination. *Adv. Funct. Mater.* **27**, 1604468 (2017).

47. Wang, Q. H., Kalantar-Zadeh, K., Kis, A., Coleman, J. N. & Strano, M. S. Electronics and optoelectronics of two-dimensional transition metal dichalcogenides. *Nat. Nanotechnol*. **7**, 699-712 (2012).

48. Zhang, H. et al. Optical thickness identification of transition metal dichalcogenide nanosheets on transparent substrates. *Nanotechnology* **28**, 164001 (2017).

49. Taghavi, N. S. et al. Thickness determination of MoS$_2$, MoSe$_2$, WS$_2$ and WSe$_2$ on transparent stamps used for deterministic transfer of 2D materials. *Nano Res*. **12**, 1691-1695 (2019).

50. Casiraghi, C. et al. Rayleigh imaging of graphene and graphene layers. *Nano Lett.* **7**, 2711-2717 (2007).

51. Li, Y. et al. Accurate identification of layer number for few-layer WS$_2$ and WSe$_2$ via spectroscopic study. *Nanotechnology* **29**, 124001 (2018).

52. Li, Y. et al. Probing symmetry properties of few-layer MoS$_2$ and h-BN by optical second-harmonic generation. *Nano Lett*. **13**, 3329-3333 (2013).

53. Yang, Y. et al. Brittle fracture of 2D MoSe$_2$. *Adv. Mater.* **29**, 1604201 (2017).

54. Nečas, D. & Klapetek, P. Gwyddion: an open-source software for SPM data analysis. *Cent. Eur. J. Phys.* **10**, 181-188 (2012).